\documentclass[showpacs,notitlepage,twocolumn,superscriptaddress,nofootinbib,preprintnumbers]{revtex4-2}
\usepackage{graphicx} 
\usepackage{bbold} 
\usepackage{tcolorbox}
\usepackage[colorlinks=true,linkcolor=blue,citecolor=blue,urlcolor=blue]{hyperref}
\usepackage{algorithm}
\usepackage{algpseudocode}
\usepackage{amsmath}
\usepackage{amsfonts}
\usepackage{amsthm}
\usepackage{amssymb}
\usepackage{mathtools}
\usepackage{braket}
\theoremstyle{plain}
\newtheorem{theorem}{Theorem}

\newtheorem{corollary}[theorem]{Corollary}

\theoremstyle{definition}

\theoremstyle{remark}

\begin{document}
\title{Toward Foundation Models for Quantum Dynamics with Provable Generalization}

\author{Guijing Duan}
\affiliation{Department of Physics, Tsinghua University, Beijing 100084, China}

\author{Chen Mo}
\affiliation{Department of Physics, Tsinghua University, Beijing 100084, China}

\author{Di Luo}
\email{diluo@tsinghua.edu.cn}
\affiliation{Department of Physics, Tsinghua University, Beijing 100084, China}
\affiliation{Institute of Advanced Study, Tsinghua University, Beijing 100084, China}
\date{October 2026}
\begin{abstract}
Understanding quantum many-body dynamics is central to uncovering nonequilibrium phenomena and advancing quantum simulation.
However, efficiently characterizing observable evolution across initial states, Hamiltonian parameters, and evolution times remains a major challenge. 
Here, we develop a foundation model built on physical principles that learns observable dynamics from noisy measurements without requiring knowledge of the microscopic form of the Hamiltonian.
For every Hamiltonian with finite-range, static or smoothly time-dependent interactions and a fixed number of tunable parameters, we prove that observable dynamics can be learned with a sample complexity independent of system size and polynomial in the evolution horizon.
Although trained only on random product states, the model generalizes to locally flat test distributions that include entangled states, and a single set of classical-shadow measurements predicts a broad class of observables at a cost logarithmic in system size.
Numerical simulations of static and periodically driven transverse-field Ising models, together with experimental data from a superconducting quantum processor, confirm accurate predictions at unmeasured settings, including coherence-sensitive dynamics. 
Our results opens up a promising path toward foundation models for quantum many-body dynamics with provable generalization.
 
\end{abstract}
\maketitle

\section{Introduction}
The real-time dynamics of quantum many-body systems underlie a broad range of phenomena, including transport, thermalization, and entanglement growth~\cite{Bertini2021,Polkovnikov2011,Andersen2025}. In regimes where such dynamics are intractable to simulate classically, quantum simulators provide a powerful platform for their experimental study~\cite{Georgescu2014,GrossBloch2017,Chen2025,Shaw2024,Shapira2025}. Systematically characterizing observable dynamics, however, requires exploring a high-dimensional space spanned by initial states, Hamiltonian parameters, and evolution times. Probing each point in this space entails a separate round of state preparation and time evolution, typically followed by many repeated measurements to estimate expectation values accurately. Consequently, exhaustive experimental exploration quickly becomes prohibitively demanding.

Recent work has increasingly focused on generalizable learning in quantum many-body physics.
For ground states, both provably sample-efficient learning algorithms~\cite{duan2609,huang2022provably,lewis2024improved} and foundation neural quantum states~\cite{RendeEtAl2025FNQS,ZaklamaEtAl2026Attention,ZhangJiangLuo2026FNEH,SinibaldiEtAl2026Designing} have been developed to predict physical properties across Hamiltonian families.
For quantum dynamics, learning-theoretic studies have established provable sample-complexity guarantees for predicting quantum processes and observable evolution, including out-of-distribution generalization beyond the training samples~\cite{HuangChenPreskill,Chen_arxiv_2409.03684,CaroEtAl2023OOD,
LiEtAl2025ChannelShadows,ZhaoEtAl2024GateComplexity,
JerbiEtAl2026Incoherent,ZhangEtAl2024Compilation,
MolteniEtAl2026Advantages,DuEtAl2025Circuits,
liao2026demonstration,BartheEtAl2025Fourier,
VallsEtAl2026Forecasting}, though these results typically address a fixed dynamical map or generalize along only a subset of initial states, Hamiltonian parameters, and evolution times.
In parallel, data-driven forecasting methods and neural-network representations of quantum states, operators, and propagators have been explored to capture quantum evolution across varied initial conditions and Hamiltonian parameters~\cite{
ErpenbeckEtAl2026ESPRIT,KanekoEtAl2025DMD,YinEtAl2023DMD,
HuangEtAl2025CNN,WuEtAl2021Forecasting,LinEtAl2021LSTM,
ZengEtAl2024Nonadiabatic,ChoiEtAl2022QNODE,Xie2026,
wang2024separable,du2026artificial,
MazzaEtAl2021Generators,HerreraRodriguez2021CNN,
NelsonEtAl2022Propagation,HuangEtAl2026Semiclassical,
ZengEtAl2026Propagators,ZhangChen2025NonMarkovian,
ZhangEtAl2024Universal,PatelEtAl2025OperatorLearning,
SinghEtAl2024Wavepacket,SecorEtAl2021Propagators,
YaoEtAl2023Distillation,ShahCangi2026TDDFT,
ZhangEtAl2025Driven,ShahEtAl2026,Mizera2023Scattering,
QiEtAl2026Floquet,QiEtAl2026NOQS,QiEtAl2026UNP}.
Yet, it remains an open challenge to establish a unified, provably generalizable framework capable of predicting observable dynamics jointly across initial states, Hamiltonian parameters, and evolution times from finite and noisy experimental measurements. 

Inspired by foundation models in artificial intelligence~\cite{BommasaniEtAl2021Foundation}, an appealing paradigm for quantum many-body dynamics is to learn, directly from experimental measurements and without a microscopic Hamiltonian model, a unified model that predicts observable expectation values at unmeasured initial states, Hamiltonian parameters, and evolution times.
Realizing this paradigm across an entire dynamical family is challenging because even an initially local observable spreads over many sites under time evolution, while longer horizons demand finer resolution in both parameters and time.
Moreover, experimental measurements yield only noisy estimates, and the implemented Hamiltonian is known only approximately due to miscalibration, crosstalk, and other imperfections.
A complementary route first identifies the Hamiltonian or noise model of a device and then simulates it~\cite{Bairey2019,Anshu2021,huang2023learning,flammia2020efficient,van2023probabilistic}, but this requires a correctly specified model class and tractable classical simulation, offering little leverage precisely in the classically intractable regime where quantum simulators are most valuable~\cite{wang2026learnable}.
A central question, therefore, is whether local quantum dynamics admits a shared representation  that can be reliably extracted from finite, noisy data without a microscopic model of the Hamiltonian, and what fundamental principles govern the cost to build such foundation model.

\begin{figure*}[t]
    \centering
    \includegraphics[width=1.0\textwidth]{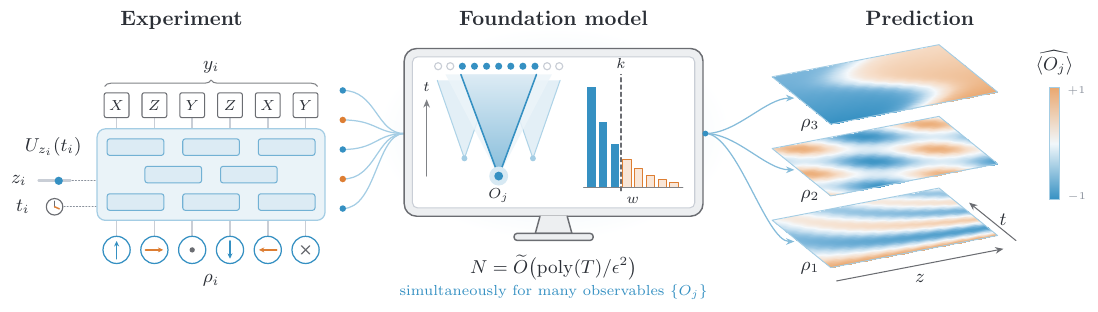}
    \caption{
Learning observable dynamics from quantum data. Measurements of evolved states provide noisy estimates of the expectation values of  observables $O_j$ for different initial states $\rho$, Hamiltonian parameters $z$, and evolution times $t$. These samples train a foundation model that exploits the effective light cone of $O_j$ to restrict the relevant spatial region and retains only Pauli strings up to weight cutoff $k$, achieving a sample complexity independent of system size. Once trained, the model estimates the expectation value for a new input without another evolution experiment, provided that the required initial-state Pauli expectation values are available. 
}
    \label{fig:0}
\end{figure*}
 
In this work, we develop a foundation model built on physical principles that learns directly from noisy measurements and generalizes jointly across initial states $\rho$, Hamiltonian parameters $z$, and evolution times $t \in [0,T]$ (Fig.~\ref{fig:0}), using only the data and the lattice geometry without any microscopic form of the Hamiltonian.
For every Hamiltonian with finite-range, static or smoothly time-dependent interactions and a fixed number of tunable parameters, we prove that the sample complexity for a bounded-degree observable is independent of system size and scales as $\widetilde O(\mathrm{poly}(T)/\varepsilon^2)$ at fixed Pauli-weight cutoff.
A single dataset of classical shadow measurements predicts a wide class of observables simultaneously, with only a logarithmic overhead in system size.
Numerical benchmarks on static and driven transverse-field Ising chains, including entangled and coherence-sensitive initial states, and experimental data from a superconducting quantum processor confirm accurate predictions at unseen settings.

\section{Learning a Family of Quantum Dynamics}
\label{sec:learning-family}
In this section, we formulate the learning problem and establish theoretical guarantees for predicting observable dynamics from finite, noisy experimental measurements.
Consider an $n$-qubit system initialized in a state $\rho$ and evolving under a Hamiltonian $H(z,\tau)$, where $z$ denotes experimentally tunable parameters and $\tau$ represents time. Experiments probe these dynamics through observables such as local magnetizations, densities, and correlation functions. For a prescribed observable $O$, the expectation value at time $t$ is given by
\begin{equation}
    F_O(z,t,\rho)
    =
    \operatorname{tr}\!\left[
        O U_z(t,0)\rho U_z(t,0)^\dagger
    \right],
    \label{eq:prediction-target}
\end{equation}
where $U_z(t,0)=\mathcal{T}\exp\!\left[-i\int_0^t H(z,\tau)\,d\tau\right]$ and $\mathcal{T}$ denotes time ordering. Varying the initial state $\rho$, parameters $z$, and evolution time $t$ generates a continuous family of expectation values. Learning the underlying structure of this family enables accurate predictions across unseen settings from a limited set of measurements.

Specifically, we consider Hamiltonians on a $d$-dimensional lattice of the form
\begin{equation}
    H(z,\tau)=H_0(\tau)+\sum_{j=1}^{m_0}z_jV_j(\tau),
    \label{eq:hamiltonian-family}
\end{equation}
where $z\in[-1,1]^{m_0}$ are tunable parameters and $m_0$ is independent of $n$.
The parameters $z_j$ may represent global fields, coupling strengths, or the amplitudes of driving protocols,  naturally accommodating both static and time-dependent dynamics.
The operators $H_0(\tau)$ and $V_j(\tau)$ are not available to the learner.
We assume only that the interactions are finite-range, have bounded strength per site, and depend smoothly on time; their microscopic form, including the specific operators $H_0(\tau)$ and $V_j(\tau)$, need not be known.

We consider the  evolution times $0\leq t\leq T$ and normalized observables $O$ satisfying $\|O\|\leq 1$  that decompose into sums of few-body terms, where each qubit participates in at most a constant number of terms. Such observables are referred to as \emph{bounded-degree}~\cite{HuangChenPreskill}.

The training data are generated by evolving randomly prepared product states under varied Hamiltonian parameters and evolution times.
Initial states are sampled from a product distribution $\mathcal D_0$ over product states $\rho=\bigotimes_{j=1}^{n}|s_j\rangle\langle s_j|$, where each $|s_j\rangle$ is drawn independently and uniformly from the six Pauli eigenstates $\{|\pm x\rangle,|\pm y\rangle,|\pm z\rangle\}$.
For each initial state, we independently draw $(z,t)\sim\nu$, where $\nu$ is a distribution over $[-1,1]^{m_0}\times[0,T]$.
After evolving the state to time $t$, we measure each qubit in a randomly selected Pauli basis.
Each experiment thus yields a single randomized measurement record, known as a classical-shadow snapshot~\cite{Huang2020}, from which unbiased estimates of the expectation values of $O$ and its local components can be computed and used as training labels.

Crucially, training on these simple product states does not restrict predictions to product states alone. Our theoretical guarantees naturally extend to any test state distribution satisfying \emph{local flatness}~\cite{HuangChenPreskill}, defined by invariance under independent single-qubit Clifford rotations. This property enables our model to learn observable dynamics from experimentally accessible product states while generalizing to a much broader class of initial states, including entangled states.

We now establish our main theoretical guarantee.
In the Heisenberg picture, the evolved observable can be expanded in Pauli strings, and the weight of a string is the number of qubits on which it acts nontrivially.
Its expectation value probes a correlation among that many qubits of the initial state, and such high-order correlations contribute exponentially little on average over the initial states considered here (Sec.~\ref{subsec:low-weight}).
We therefore retain only strings of weight at most $k$, which incurs a truncation error $\Delta_k$ that decays exponentially with $k$.
The total prediction error then decomposes into the truncation error $\Delta_k$ and a learning error $\varepsilon$ controlled by the number of training samples.

\begin{theorem}[Learning a dynamical family]
\label{thm:joint-dynamical-learning}
Consider the $m_0$-parameter Hamiltonian family $H(z,\tau)$ on a
$d$-dimensional lattice of $n$ qubits with evolution horizon $T$ and  product
distribution $\mathcal D_0$ specified above, and fix a normalized
bounded-degree observable $O$ before collecting data.
For any Pauli weight cutoff $k\geq 0$, target accuracy $0<\varepsilon\leq1$,
and failure probability $0<\delta<1/2$,
there exists a model $\widehat F_O$ 
that can be learned from $N_{\mathrm{fixed}}$
independent training samples, each containing
a single classical shadow snapshot, such that,
with probability at least $1-\delta$,
\begin{equation}
    \left(
        \mathbb E_{\substack{(z,t)\sim\nu\\ \rho\sim\mathcal D}}
        \left|
            \widehat F_O(z,t,\rho)-F_O(z,t,\rho)
        \right|^2
    \right)^{1/2}
    \leq  \Delta_k+\varepsilon
    \label{eq:main-prediction-guarantee}
\end{equation}
for every locally flat test state distribution
$\mathcal D$.
Here, the high-weight truncation error
is bounded by
$\Delta_k=\|O\|(2/3)^{(k+1)/2}$, which tightens to
$\Delta_k=\|O\|3^{-(k+1)/2}$ when $\mathcal D = \mathcal D_0$.
The required number of experimental samples satisfies
\begin{equation}
    N_{\mathrm{fixed}}
    =
    \widetilde O\left[
        \frac{(1+T)^{\alpha}}{\varepsilon^2}
        \log\frac{1}{\delta}
    \right]
    \label{eq:main-sample-complexity}
\end{equation}
where $\alpha=dk+(d+1)m_0+1$.
\end{theorem}

For a fixed cutoff $k$ and target accuracy $\varepsilon$, the sample complexity grows at most polynomially with the evolution horizon $T$ and remains independent of the system size $n$.
This guarantee holds equally for static and smoothly time-dependent Hamiltonians, with the full proof provided in Appendices~B--E.

Moreover, the same framework naturally specializes to settings in which a subset of the inputs $(\rho, z, t)$ is held fixed, simply by dropping the model's dependence on the fixed inputs.
In particular, when the initial state $\rho$ is fixed and one generalizes across Hamiltonian parameters $z$ and evolution times $t$, the Pauli-weight truncation error $\Delta_k$ vanishes and the $dk$ term drops out, reducing the horizon exponent to $\alpha = (d+1)m_0 + 1$.
Further fixing one of the remaining inputs yields two special cases: (\textbf{i})~for purely temporal generalization (fixed $z$), the $(d+1)m_0$ term drops out, giving a scaling linear in $T$ ($\alpha = 1$) that is unconditionally worst-case optimal up to logarithmic factors; and (\textbf{ii})~for parameter generalization at a fixed readout time (fixed $t$), the $+1$ term drops out, giving an exponent of $\alpha = (d+1)m_0$.

Beyond predicting a single observable, the same experimental dataset can be directly reused to predict an entire family of observables simultaneously.

\begin{corollary}[Simultaneous prediction of observables]
\label{cor:uniform-observables}
Under the assumptions of Theorem~\ref{thm:joint-dynamical-learning},
consider the class of all normalized bounded-degree
observables with fixed locality and degree bounds.
Given a single training dataset of size
\begin{equation}
    N_{\mathrm{all}}
    =
    \widetilde O\left[
        \frac{(1+T)^{\alpha}}{\varepsilon^2}
        \log\frac{n}{\delta}
    \right],
    \label{eq:uniform-sample-complexity}
\end{equation}
with  $\alpha=dk+(d+1)m_0+1$ and probability at least $1-\delta$, the resulting
models $\{\widehat F_O\}$ satisfy Eq.~\eqref{eq:main-prediction-guarantee}
simultaneously for all observables $O$ in this class
and all locally flat test state distributions $\mathcal D$.
\end{corollary}

In practice, this means that a single set of experimental measurements can be reused to learn arbitrary observables in the class, requiring only additional classical post-processing rather than new evolution experiments. Remarkably, extending the theoretical guarantee to the entire observable class incurs only a logarithmic overhead in the system size $n$.

\section{Physical Origin of Learnability}
\label{sec:III}
We now explain the physical mechanisms behind Theorem~\ref{thm:joint-dynamical-learning}.
The expectation value of an observable can exhibit markedly different dynamics across varied initial states, even under the same Hamiltonian.
In the Heisenberg picture, however, all such trajectories are governed by a common time-dependent operator,
\begin{equation}
    F_O(z,t,\rho)
    =
    \operatorname{tr}\!\left[\rho\,O(z,t)\right],
    \quad
    O(z,t)
    =
    U_z(t,0)^\dagger O U_z(t,0).
    \label{eq:heisenberg-response}
\end{equation}
Because the Heisenberg operator $O(z,t)$ depends only on the Hamiltonian parameters $z$ and evolution time $t$, it encodes all dynamical information independently of the initial state, whereas $\rho$ enters purely through the linear overlap $\operatorname{tr}[\rho O(z,t)]$.
This separation decouples the dynamical evolution from the choice of initial state, so that once the relevant components of $O(z,t)$ are learned from measurements on simple product states, the same operator representation predicts expectation values for arbitrary initial states, including highly entangled ones, solely through their initial correlations.
At the same time, it allows us to analyze the intrinsic locality and smoothness of $O(z,t)$ independently of any particular state, while using the initial-state distribution to identify which operator components contribute appreciably to the prediction error.

The learnability of this dynamical family relies on three key properties. First, locality restricts the spatial spreading of initially local observables to an effective light cone. Second, across locally flat initial state distributions, the contribution of high-weight Pauli strings to the average prediction error is exponentially suppressed, justifying a low-weight approximation. Third, the smooth dependence of the remaining Pauli coefficients on Hamiltonian parameters and evolution time can be captured by a finite set of basis functions.  Together, these three mechanisms yield a foundation model of observable dynamics across the entire family, which we examine in turn below.

\subsection{Locality limits the relevant spatial region}

\label{subsec:locality}

A bounded-degree observable may contain terms
distributed throughout the system, although each
term acts on only a constant number of qubits.
Writing $O=\sum_\alpha O_\alpha$, the linearity
of Heisenberg evolution gives
$O(z,t)=\sum_\alpha O_\alpha(z,t)$.
We can therefore understand the evolution of
$O$ by examining how its individual terms spread.

For finite-range interactions, the spatial spreading of
a local operator is constrained by locality.
Starting from its initial support, the operator
can propagate to distant sites only through
successive local interactions, as illustrated in Fig.~\ref{fig:0}.
At any given evolution time, nontrivial contributions are
largely confined within an effective light cone
whose radius expands linearly with time, whereas
contributions outside this cone are exponentially
suppressed~\cite{LiebRobinson1972,Barthel2012}.
The associated light cone velocity is determined by the
interaction range and per-site interaction strength, remaining
bounded independently of the total system size.
 
This spatial restriction can be formalized
by considering the evolution within a finite
neighborhood of the initial operator.
Let $S_{O_\alpha}(R)$ denote the region containing all sites within
distance $R$ of the support of $O_\alpha$,
and let $O_{\alpha,R}(z,t)$ denote the operator evolved
under the Hamiltonian restricted to interactions
supported entirely within $S_{O_\alpha}(R)$.
The Lieb--Robinson bound gives
\begin{equation}
    \bigl\|O_\alpha(z,t)-O_{\alpha,R}(z,t)\bigr\|
    \leq
    C\|O_\alpha\|e^{-\mu(R-v|t|)},
    \label{eq:locality-window-bound}
\end{equation}
where the bounded support size of $O_\alpha$
is absorbed into $C$, and the constants $C$, $\mu$, and $v$ can be chosen
uniformly across the Hamiltonian family and
independently of the system size.
For evolution times up to $T$, choosing $R = vT + \mu^{-1}\log(C/\varepsilon)$ ensures that the truncation error in Eq.~\eqref{eq:locality-window-bound} is at most $\varepsilon\|O_\alpha\|$, so that the resulting expectation value depends only on the reduced density matrix of the initial state on $S_{O_\alpha}(R)$ up to error $\varepsilon\|O_\alpha\|$.
Because the error in Eq.~\eqref{eq:locality-window-bound} is bounded in the operator norm, this spatial truncation holds uniformly for all initial states.
Consequently, the relevant neighborhood for each local term is independent of the system size and contains $\ell_R = O\bigl((1 + vT + \log(1/\varepsilon))^d\bigr)$ sites.

Although locality substantially restricts the spatial
region required to evaluate an observable, it
does not immediately guarantee an efficient representation.
Even within the light cone, an evolving operator
can contain exponentially many Pauli strings.
Identifying which of these components contribute
appreciably to observable expectation values is therefore
crucial for constructing a tractable model
of quantum dynamics.

\subsection{Limited sensitivity to high-order correlations}
\label{subsec:low-weight}

Even within the effective light cone, local interactions can generate Pauli strings acting on an increasing number of qubits.
Nevertheless, the extent to which these complex components affect observable expectation values depends on the initial state distribution.
For suitable distributions, accurate predictions can require far less information than a full reconstruction of the evolving operator.

To see this, we expand the localized observable $O_{\alpha,R}(z,t)$
in the Pauli basis,
\begin{equation}
    O_{\alpha,R}(z,t)
    =
    \sum_{P\in \{I,X,Y,Z\}^{\otimes n}} b_{\alpha,P}(z,t)P.
    \label{eq:localized-pauli-expansion}
\end{equation}
A Pauli string $P$ of weight $w$ acts nontrivially
on $w$ qubits, so its expectation value
measures a $w$-body correlation in the initial
state. For the product-state training distribution
$\mathcal D_0$, this expectation value is
nonzero only when the preparation basis of
each of the $w$ active qubits aligns with the corresponding
Pauli operator in $P$.
Because each qubit basis is drawn independently and uniformly
from $X$, $Y$, and $Z$, this alignment occurs with probability $3^{-w}$.
Consequently, the mean squared expectation value over $\mathcal D_0$
is exactly $3^{-w}$, meaning that
high-weight Pauli strings are exponentially
suppressed across the training distribution.

This exponential suppression generalizes beyond product states to any initial state distribution satisfying local flatness.
For any locally flat distribution $\mathcal D$,
the average squared expectation value of a
weight-$w$ Pauli string is bounded by
$(2/3)^w$ \cite{HuangChenPreskill}.
Moreover, local flatness eliminates cross terms
between distinct Pauli strings when averaging the squared prediction error over $\mathcal D$.
Each Pauli component therefore contributes additively to the
mean squared prediction error, weighted
by its mean squared expectation value over
the initial state distribution.

At the same time, unitary evolution constrains
the growth of the Pauli coefficients themselves.
By unitarity, the total squared magnitude
of the Pauli coefficients is conserved over time, satisfying
$\sum_P |b_{\alpha,P}(z,t)|^2
=\sum_P |b_{\alpha,P}(z,0)|^2
\leq \|O_\alpha\|^2$.
Thus, while local interactions redistribute spectral
weight across increasingly complex
Pauli strings, they cannot amplify the total weight.
Combining this conservation law with the exponential suppression of high-weight correlations directly bounds the prediction error incurred by truncating high-weight Pauli strings.

Let $O_{\alpha,R}^{(\leq k)}(z,t)$ denote
the operator obtained by retaining only strings
of weight at most $k$ within the radius-$R$ light-cone region.
For any locally flat distribution $\mathcal D$,
\begin{equation}
\begin{aligned}
\Biggl(
\mathbb E_{\rho\sim\mathcal D}
\Bigl|
\operatorname{tr}\Bigl[
\rho\Bigl(
O_{\alpha,R}(z,t)
&-
O_{\alpha,R}^{(\leq k)}(z,t)
\Bigr)
\Bigr]
\Bigr|^2
\Biggr)^{1/2}
\\
&\leq
\|O_\alpha\|
\left(\frac{2}{3}\right)^{(k+1)/2}.
\end{aligned}
\label{eq:low-weight-prediction-bound}
\end{equation}
For the training distribution $\mathcal D_0$,
the factor on the right-hand side is instead
$3^{-(k+1)/2}$.
Crucially, these bounds hold uniformly across all
Hamiltonian parameters and evolution times.
 
Together, locality and high-weight truncation
substantially reduce the number of Pauli strings
needed for prediction.
For a light-cone neighborhood containing $\ell_R$
qubits, retaining strings of weight at most $k$
leaves only $O(\ell_R^k)$ components with fixed exponents $k$,
rather than exponentially many.

\subsection{Finite representation across parameters and time}
\label{subsec:parameter-time-dependence}

Having identified the Pauli components relevant
for prediction, we now examine how their coefficients
$b_{\alpha,P}(z,t)$ depend on the Hamiltonian parameters and evolution time.
As $z$ and $t$ vary, the accumulation of quantum phases
and the interference among different dynamical processes
can induce intricate oscillations in $b_{\alpha,P}(z,t)$.
Despite these oscillations, the underlying dependence on the parameters and time remains smooth, allowing a controlled approximation
within a finite basis expansion.

 This representation applies broadly across static and time-dependent Hamiltonian families, without requiring integrability, translation invariance, periodic driving, or knowledge of the microscopic interaction terms.
Specifically, we expand each retained Pauli coefficient in products of Chebyshev polynomials,
\begin{equation}
b_{\alpha,P}(z,t)
\approx
\sum_{\lambda=1}^{M}
c_{\alpha,P,\lambda}\phi_\lambda(z,t),
\end{equation}
where $\phi_\lambda(z,t)$ are multivariate basis functions of the Hamiltonian parameters and rescaled evolution time.
Under our locality, bounded-strength, and real-analyticity assumptions, each coefficient $b_{\alpha,P}(z,t)$ extends analytically to a complex neighborhood of $[-1,1]^{m_0}\times[0,T]$, ensuring that the Chebyshev truncation error decays exponentially with the polynomial degrees.

\section{Building a Foundation Model for Quantum Dynamics}
\label{sec:learning-algorithm}
\subsection{Model construction and training}
We first describe the learning procedure for
an observable $O$ consisting of a single Pauli string acting
on a bounded number of qubits, such as $X_i$, $Z_iZ_j$, or $X_iY_jZ_k$.
General bounded-degree observables can be treated
as linear combinations of such local Pauli strings.
For each independently sampled input
$x_i=(\rho_i, z_i,t_i)$, let
$\widehat\sigma_{i,1},\ldots,\widehat\sigma_{i,M}$
denote the classical-shadow snapshots obtained
from repeated preparations and measurements \cite{Huang2020}.
They provide a training label
\begin{equation}
    \overline Y_{i,O}
    =
    \frac{1}{M_s}
    \sum_{m=1}^{M_s}
    \operatorname{tr}\!\left(
        O\widehat\sigma_{i,m}
    \right).
    \label{eq:algorithm-observable-label}
\end{equation}
Although averaging multiple snapshots reduces
measurement noise, a single snapshot per training
sample is sufficient for learning with theoretical guarantees.
Shot noise is suppressed statistically across the ensemble of
independent training inputs, eliminating the need for
precise expectation value estimates at any
individual setting.

The finite basis representation developed in Sec.~\ref{subsec:parameter-time-dependence} directly informs our model architecture.
For a local Pauli observable $O$, we retain
Pauli strings within $S_O(R)$ of weight at
most $k$, and expand their parameter- and time-dependent coefficients
in Chebyshev polynomials.
This yields a learning model of the form
\begin{equation}
\begin{aligned}
    &F_{O,w}(z,t,\rho)
    =
    \langle w,\phi_O(z,t,\rho)\rangle, \\
    &\phi_O(z,t,\rho)
    =
    \left(
        \operatorname{tr}(P\rho)\,
        \psi_{\boldsymbol r}(z)\,
        \psi_q\!\left(\frac{2t}{T}-1\right)
    \right)_{P,\boldsymbol r,q}.
        \end{aligned}
    \label{eq:algorithm-input-features}
\end{equation}
Here, $\psi_m$ denotes the normalized $m$-th Chebyshev polynomial of the first kind, $\boldsymbol r = (r_1,\dots,r_{m_0})$ is a multi-index with $\psi_{\boldsymbol r}(z)=\prod_{j=1}^{m_0}\psi_{r_j}(z_j)$, and $p_z$ and $p_t$ are the polynomial truncation degrees in the Hamiltonian parameters and evolution time, respectively, with $0\leq r_j\leq p_z$ and $0\leq q\leq p_t$.
The index $P$ runs over all Pauli strings
supported within $S_O(R)$ with weight
$|P|\leq k$.
For the product states used in training,
$\operatorname{tr}(P\rho)$ can be calculated
directly from the single-qubit states
prepared in each sample.
Each input $(z,t,\rho)$ thus specifies a known feature vector $\phi_O(z,t,\rho)$, whereas
the learnable weight vector $w$ encodes information of the underlying
quantum dynamics.

We estimate the weights $w$ from the noisy labels $\overline Y_{i,O}$
via ridge regression:
\begin{equation}
    \widehat w_O
    =
    \underset{w}{\operatorname{argmin}}
    \left\{
        \frac{1}{N}
        \sum_{i=1}^{N}
        \left[
            \overline Y_{i,O}
            -
            F_{O,w}(z_i,t_i,\rho_i)
        \right]^2
        +
        \lambda\|w\|_2^2
    \right\},
    \label{eq:algorithm-training}
\end{equation}
with regularization strength $\lambda>0$.
The first term fits the measured expectation
values, while the second stabilizes the fit
against measurement noise by penalizing
large coefficients.
Eq.~\eqref{eq:algorithm-training} is a convex quadratic program with a closed-form solution.
The explicit choices of truncation degrees and regularization strength
that guarantee the error bound in
Eq.~\eqref{eq:main-prediction-guarantee}
are detailed in Appendix~D.

After training, the model $\widehat F_O=F_{O,\widehat w_O}$ predicts expectation values across new initial states, Hamiltonian parameters, and evolution times without retraining.
Crucially, although $\widehat w_O$ is learned exclusively from random product states, the model generalizes to broader locally flat test distributions that include entangled initial states.
For such states, the multi-qubit features $\operatorname{tr}(P\rho)$ with $2 \leq |P| \leq k$ capture nonfactorizable initial correlations and quantum coherence that directly contribute to the evolved observable.
Moreover, prediction does not require a full classical description of $\rho$. Even for an unknown initial state drawn from a locally flat distribution, knowing or measuring only its low-weight Pauli expectation values $\operatorname{tr}(P\rho)$ within $S_O(R)$ suffices to evaluate $\widehat F_O$ without performing any additional time-evolution experiments.

The same construction extends to any
bounded-degree observable
$O=a_I I+\sum_{Q\in\mathcal{Q_O}}a_QQ$,
where $\mathcal Q_O$ collects the non-identity Pauli strings.
We train a separate component model $\widehat F_Q$ for each $Q\in\mathcal Q_O$
using labels $\overline Y_{i,Q}$ evaluated
on the same classical shadow dataset.
Together, the learned local component models $\{\widehat F_Q\}$ form a reusable foundation model for the dynamical family, from which the full observable prediction is assembled by linearity as
\begin{equation}
    \widehat F_O(z,t,\rho)
    =
    a_I+
    \sum_{Q\in\mathcal Q_O}
    a_Q\,\widehat F_Q(z,t,\rho).
    \label{eq:algorithm-combined-prediction}
\end{equation}
Since the identity term $a_I$ is known trivially and
$\sum_{Q\in\mathcal Q_O}|a_Q|$ is bounded by a system-size-independent constant for any normalized bounded-degree observable, the combined prediction
error remains controlled.

Learning the Pauli components separately allows
their models to be reused for different
observables.
For example, predictions for $X_iX_j$,
$Y_iY_j$, and $Z_iZ_j$ can be combined to
obtain the bond energy
$\mathbb{h}_{ij}=J_\perp(X_iX_j+Y_iY_j)+J_zZ_iZ_j$,
or other observables formed from the same
components.
Alternatively, one can linearly combine the shadow labels
prior to optimization and train a single model
directly for a target composite observable,
an equivalence we analyze in Appendix~F.

Finally, we distinguish between classical and experimental resource requirements.
For a fixed weight cutoff $k$ and parameter dimension $m_0$,
the feature dimension of each component model scales at most
polynomially with the evolution horizon $T$.
Although a spatially extensive observable comprises $O(n)$ local terms
and thus entails a classical storage and post-processing cost that grows with system size,
the required number of quantum experimental samples remains independent of $n$ for any fixed observable.
Bounding the estimation error uniformly across all local Pauli components
yields the logarithmic $n$-dependence in Eq.~\eqref{eq:uniform-sample-complexity}.
 \begin{figure*}[!t]
    \centering
    \includegraphics[width=1.0\textwidth]{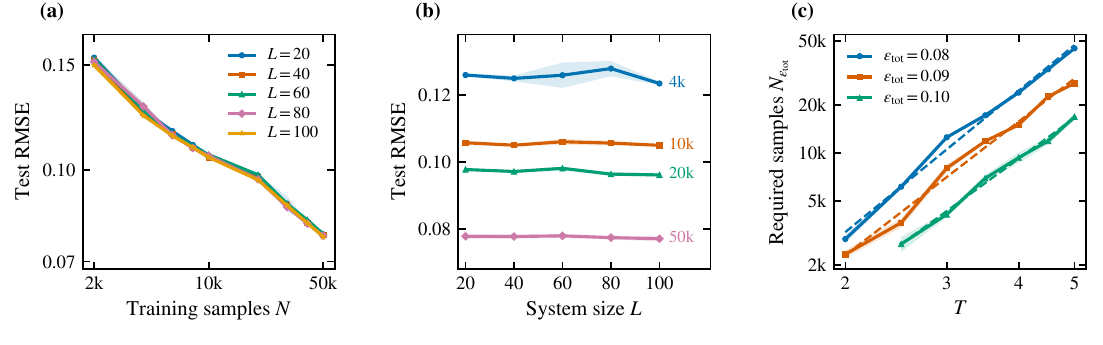}
    \caption{
Learning local magnetization dynamics in the one-dimensional transverse-field Ising model.
(a)~Test root-mean-square error (RMSE) versus the number of training samples for different system sizes, where the test RMSE measures the total prediction error including both the Pauli-weight truncation error and the learning error.
(b)~Test RMSE versus system size at fixed training-set sizes.
(c)~Number of training samples $N_\varepsilon$ required to reach target total test RMSEs $\varepsilon_{\mathrm{tot}} = 0.08$, $0.09$, and $0.10$ as a function of the evolution horizon $T$.
}
    \label{fig:2}
\end{figure*}

\subsection{Proof sketch of Theorem~\ref{thm:joint-dynamical-learning} and Corollary~\ref{cor:uniform-observables}}

The exponent $\alpha = dk + (d+1)m_0 + 1$ in Theorem.~\ref{thm:joint-dynamical-learning} follows from three factors in the feature dimension.
For a local Pauli observable, the Lieb--Robinson bound restricts the relevant dynamics, up to a controlled error, to a light-cone neighborhood containing $\ell_R = \widetilde{O}((1+T)^d)$ sites.
Retaining Pauli strings of weight at most $k$ within this neighborhood leaves $O(\ell_R^k)$ initial-state correlation features, giving the  exponent $dk$.

Within the light cone, varying a Hamiltonian parameter changes
local interaction terms whose total strength is at most $O(\ell_R)$.
Their effect accumulates over the evolution time, giving
the scale $T\ell_R$ and a sufficient polynomial degree
$p_z=\widetilde{O}((1+T)^{d+1})$ for each parameter.
The tensor-product basis for $m_0$ parameters therefore contains
$(p_z+1)^{m_0}=\widetilde{O}((1+T)^{(d+1)m_0})$
features, contributing $(d+1)m_0$ to the exponent.
Meanwhile, locality and the uniform time-analyticity assumption
ensure that the evolved local observable admits a bounded
analytic continuation to a complex-time neighborhood of fixed
width, independent of the light-cone volume.
A temporal degree $p_t=\widetilde{O}(1+T)$ therefore suffices,
contributing the final $+1$.

Multiplying these factors gives a total feature dimension
$D = \widetilde{O}((1+T)^{dk+(d+1)m_0+1})$.
For fixed $k$, the approximation and ridge-regression analysis
in Appendices~C--E shows that
$N=\widetilde{O}(D\varepsilon^{-2}\log(1/\delta))$ samples
suffice to achieve the prediction guarantee.
The sample complexity therefore inherits the same horizon
exponent $\alpha$, yielding Eq.~\eqref{eq:main-sample-complexity}.
Corollary~\ref{cor:uniform-observables} follows by reusing the same classical-shadow dataset
to learn all Pauli observables of bounded weight.
Since their number grows only polynomially with $n$, a union
bound replaces $\log(1/\delta)$ by $\log(n/\delta)$, introducing
only a logarithmic overhead in system size.

\section{Numerical and Experimental Results}
In this section, we present numerical and 
experimental results to demonstrate the effectiveness of our model.

\subsection{One-dimensional transverse-field Ising model}
We begin with a one-dimensional transverse-field Ising model (TFIM) with 
nearest-neighbor interactions, governed by the Hamiltonian
\begin{equation}
    H(h) = -\frac{1}{2}\sum_{i=1}^{L-1} Z_i Z_{i+1} - \frac{h}{2}\sum_{i=1}^{L} X_i,
    \label{eq:ising-hamiltonian}
\end{equation}  
where $X_i$ and $Z_i$ denote Pauli operators at site $i$, and $h$ is the tunable transverse field.
We denote the $Z$ eigenstates by $\ket{\uparrow}$ and $\ket{\downarrow}$,
and the $X$ eigenstates by
$\ket{\rightarrow}=(\ket{\uparrow}+\ket{\downarrow})/\sqrt2$
and $\ket{\leftarrow}=(\ket{\uparrow}-\ket{\downarrow})/\sqrt2$.
We train a foundation model to predict the expectation
value of the local magnetization $X_{L/2}$ jointly across
initial states, field strengths, and evolution times.
Each training sample is generated by independently
drawing $h$ uniformly from $[-1,1]$, $t$ uniformly
from $[0,T]$, and each qubit state uniformly from
the six Pauli eigenstates.
For each sample, the training label is obtained by adding zero-mean unbiased noise to the exact expectation value to emulate finite-shot measurement noise, matching the unbiased noisy-label model of our theoretical analysis 

Hyperparameters are selected via repeated holdout validation across three random 80:20 training--validation splits to minimize the pooled root-mean-square error (RMSE).
The resulting model is then refitted on the full training dataset and evaluated on an independent test set against the exact dynamics, so the reported test RMSE reflects the total prediction error, including both the Pauli-weight truncation error and the statistical learning error.
Full hyperparameter ranges and implementation details are provided in Appendix~G.

As shown in Fig.~\ref{fig:2}(a), the test prediction error decreases monotonically with the number of training samples, showing no sign of saturation over the explored range and confirming systematic generalization across initial states, field strengths, and evolution times.
Crucially, the learning curves for different system sizes nearly collapse onto one another, demonstrating that increasing the system size does not degrade the prediction accuracy. 
Furthermore, Fig.~\ref{fig:2}(b) demonstrates that the test error
remains nearly constant as the chain length $L$ grows from
$20$ to $100$ at any fixed training set size, directly corroborating our theoretical
guarantee that the sample complexity for a
normalized bounded-degree observable is
independent of system size.

We further examine how the sample complexity scales
with the evolution horizon $T$.
At a fixed target accuracy, the number of training
samples required as a function of $T$
exhibits an approximately linear trend
on a log--log scale, as shown in
Fig.~\ref{fig:2}(c),
confirming a polynomial (power-law) scaling of the sample
complexity with the evolution horizon.

\begin{figure*}[!t]
    \centering
    \includegraphics[width=1.0\textwidth]{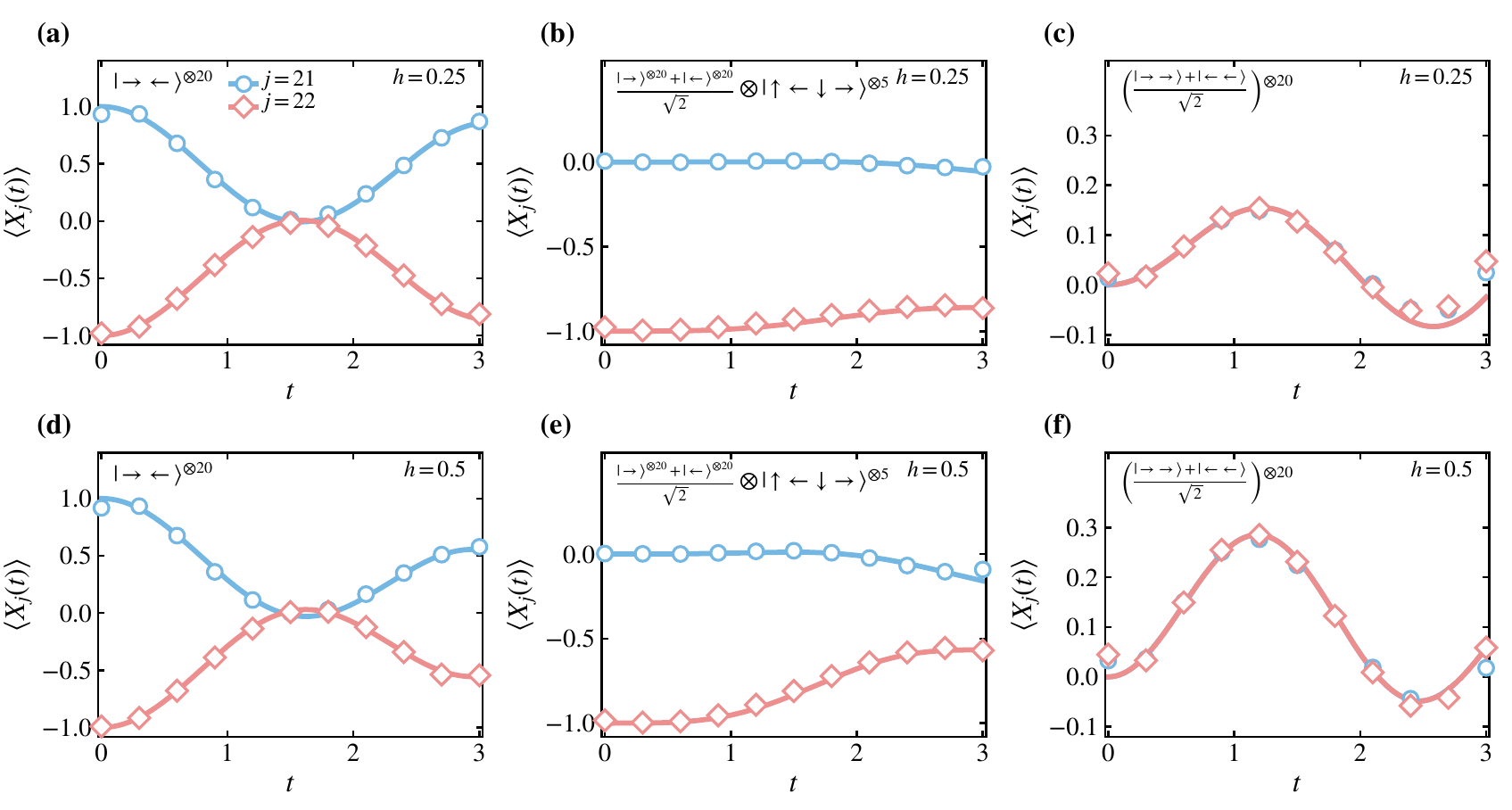}
    \caption{Predictions of  $X_j(t)$  for
different initial states in a transverse-field
Ising chain with $L=40$.
The columns correspond to (a,d) the $X$-basis N\'eel state,
(b,e) a half-chain GHZ state on the first 20 spins tensored
with a product state on the remaining spins, and
(c,f) a product of adjacent Bell pairs
$(|\rightarrow\rightarrow\rangle+
|\leftarrow\leftarrow\rangle)/\sqrt2$.
The transverse field is $h=0.25$ in the top row
and $h=0.5$ in the bottom row.
Blue and red correspond to sites $j=21$ and $j=22$,
respectively.
Solid lines show exact dynamics, and open circles
and diamonds show predictions for the two sites.
Each site-specific model is trained using
$50{,}000$ independently sampled inputs
$(\rho,h,t)$, with random Pauli product states
$\rho$. }
    \label{fig:3}
\end{figure*}

Beyond distribution-averaged test errors, we next evaluate the ability of our model to predict full time-evolution trajectories across different field strengths and continuous evolution times for specific initial states featuring distinct spatial ordering and quantum entanglement, as shown in Fig.~\ref{fig:3}.
Using $50{,}000$ random product-state samples, we train separate models for the local observables $X_{21}$ and $X_{22}$ over the time window $t \in [0, 3]$, and evaluate them directly on three representative test states at field strengths $h = 0.25$ and $h = 0.5$ without retraining.
 
First, for the spatially ordered $X$-basis N\'eel state
$|\psi_{\mathrm N}\rangle
=|\rightarrow\leftarrow\rightarrow\leftarrow\cdots\rangle$,
the predicted trajectories in Fig.~\ref{fig:3}(a,d)
closely match the exact dynamics, faithfully capturing
both the initial relaxation and subsequent revival of the
staggered local magnetizations.

We next extend the tests to entangled initial
states, which are absent from the training
distribution. We first consider a hybrid state comprising a GHZ state
on the left half of the chain and a product
state on the right,
\begin{equation}
    |\psi_{\mathrm e}\rangle
    =
    \frac{
        |\rightarrow\rangle^{\otimes m}
        +
        |\leftarrow\rangle^{\otimes m}
    }{\sqrt{2}}
    \otimes
    |\uparrow\leftarrow\downarrow\rightarrow
    \cdots\rangle,
    \quad m=L/2.
    \label{eq:entangled-test-state}
\end{equation}
As shown in Fig.~\ref{fig:3}(b,e), the models again
reproduce the exact dynamics.
However, the quantum coherence between the two GHZ
branches is encoded solely in $m$-body Pauli correlations
spanning all $m=20$ qubits of the GHZ block.
Consequently, a model with weight cutoff $k<m$ cannot
distinguish this pure state from the corresponding
incoherent mixture, and indeed the exact local
magnetization itself is insensitive to this nonlocal
coherence.

\begin{figure}[!t]
    \centering
    \includegraphics[width=1.0\columnwidth]{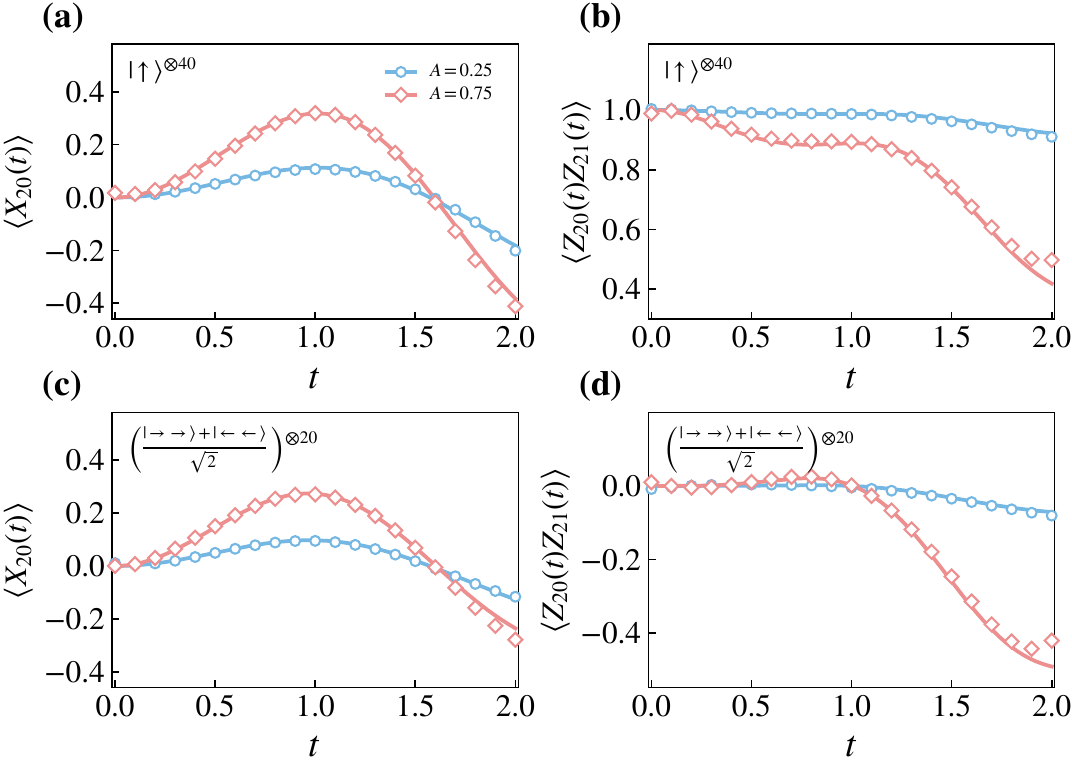}
\caption{
Learning observable dynamics under a periodically
driven transverse-field Ising Hamiltonian
for a chain of $L=40$ qubits at drive
frequency $\omega=2$.
(a,c) display
$\langle X_{20}(t)\rangle$, and (b,d)
display $\langle Z_{20}(t)Z_{21}(t)\rangle$.
The initial state is the fully polarized product
state $|\uparrow\rangle^{\otimes L}$ in (a,b)
and the Bell-pair product state
$\bigl[(|\rightarrow\rightarrow\rangle+
|\leftarrow\leftarrow\rangle)/\sqrt{2}\bigr]^{\otimes L/2}$
in (c,d).
Blue and red denote drive amplitudes
$A=0.25$ and $A=0.75$, respectively.
Solid lines show exact dynamics, and open circles
and diamonds show predictions for the two amplitudes.
Each observable model is
trained on $30{,}000$ random product state samples,
with $A$ and $t$ drawn uniformly from $[-1,1]$
and $[0,2]$, respectively.
}
    \label{fig:4}
\end{figure}

To directly probe dynamics driven by quantum entanglement, we examine a second entangled initial state
whose local observable evolution depends nontrivially on quantum coherence, namely a tensor product of adjacent Bell pairs,
\begin{equation}
    |\psi_{\mathrm B}\rangle
    =
    \bigotimes_{r=1}^{L/2}
    \frac{
        |\rightarrow\rightarrow\rangle_{2r-1,2r}
        +
        |\leftarrow\leftarrow\rangle_{2r-1,2r}
    }{\sqrt{2}}.
    \label{eq:bell-test-state}
\end{equation}
In contrast to the GHZ state, the quantum coherence within
each Bell pair generates two-qubit
Pauli correlations that directly govern $\langle X_j(t) \rangle$.
Indeed, dephasing these Bell pairs in the
$X$ basis produces an incoherent
mixture for which $\langle X_j(t)\rangle=0$
at all evolution times, whereas
the pure Bell pair state $|\psi_{\mathrm B}\rangle$
exhibits pronounced magnetization oscillations.
As demonstrated in Fig.~\ref{fig:3}(c,f), our
models accurately resolve these coherence-induced
oscillations.

These results show that models trained
exclusively on random product states
can predict the dynamics of structured
and entangled initial states, including
cases where quantum coherence directly
contributes to the observable dynamics. 

\subsection{Time-dependent Hamiltonian}

We next investigate whether our learning framework
extends to time-dependent Hamiltonians.
Starting from the transverse-field Ising model
in Eq.~\eqref{eq:ising-hamiltonian}, we replace
the static field $h$ with a periodic drive
$h(t)=A\cos(\omega t)$.
Since the instantaneous Hamiltonians at different times
generally do not commute, the evolution depends nontrivially
on the history of the applied field.
Nevertheless, the same learning framework can
be used to describe the resulting dynamics
across drive amplitudes and evolution times.

Specifically, we consider a chain of $L=40$ qubits with
$\omega=2$ and train independent models for
$X_{20}$ and $Z_{20}Z_{21}$ using $30{,}000$
random product-state samples.
The drive amplitude $A$ and evolution time $t$
are sampled uniformly from $[-1,1]$ and $[0,2]$,
respectively.

Fig.~\ref{fig:4} compares the predictions with
the exact dynamics for the fully polarized state
$\ket \uparrow^{\otimes L}$ and the Bell-pair
state defined in Eq.~\eqref{eq:bell-test-state},
at drive amplitudes $A=0.25$ and $A=0.75$.
The predictions reproduce the local operator
dynamics, including its sign reversal,
as shown in Fig.~\ref{fig:4}(a,c).
In Fig.~\ref{fig:4}(b,d), the two-site correlation exhibits distinct
behavior for the two initial states, decreasing
from unity in the polarized state while developing
negative values from zero in the Bell-pair state.
The bond $(20,21)$ connects two different Bell pairs,
so its initial $ZZ$ correlation vanishes.
These results demonstrate that our learning
approach extends to time-dependent Hamiltonians,
allowing accurate predictions of observable
dynamics for both product and entangled initial states.

\subsection{Experimental demonstration on a superconducting quantum processor}
 
To validate our approach on quantum hardware,
we apply the learning framework to experimental
measurements collected on the superconducting
quantum processor \texttt{ibm\_kingston}.
We implement a ten-qubit transverse-field
Ising chain with open boundary conditions (Eq.~\eqref{eq:ising-hamiltonian}).
For each sample, an initial state is prepared
using the same sampling procedure as in the
numerical study and evolved through a
Suzuki--Trotter decomposition \cite{Suzuki1976}, with $h$ and $t$
sampled uniformly from $[0,1]$ and $[0,1]$,
respectively.  
Measuring all qubits in the $X$ basis yields simultaneous
estimates of the ten local magnetizations
$\langle X_j\rangle$ using $1{,}024$ shots per circuit,
with full circuit details provided in Appendix~H.
Then we train the model on $1{,}000$ experimental
samples, selecting hyperparameters through
three repeated $80{:}20$ training--validation
splits within this dataset.
 
\begin{figure*}[!t]
    \centering
    \includegraphics[width=1.0\textwidth]{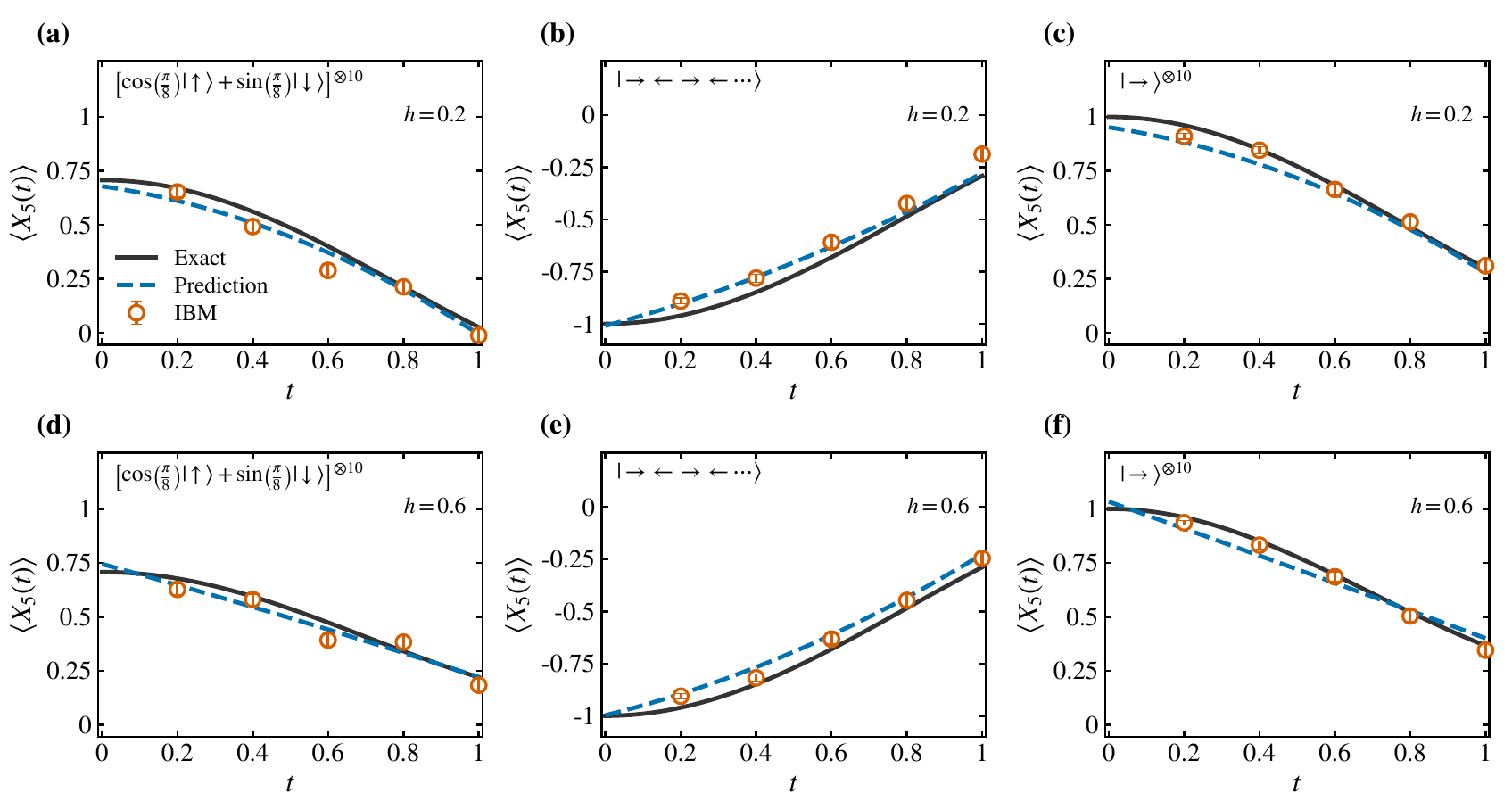}
\caption{
Prediction of $\langle X_5(t)\rangle$ dynamics
from  measurements on a superconducting quantum processor.
A model is trained on $1{,}000$ random product-state
samples collected on \texttt{ibm\_kingston}.
The test initial states are (a,d)  $[\cos(\pi/8)|\uparrow\rangle+
\sin(\pi/8)|\downarrow\rangle]^{\otimes 10}$,
(b,e) the $X$-basis N\'eel state
$|\rightarrow\leftarrow\rightarrow\leftarrow\cdots\rangle$,
and (c,f) the $X$-polarized state
$|\rightarrow\rangle^{\otimes 10}$.
The transverse field is $h=0.2$ in the top row
and $h=0.6$ in the bottom row.
Black solid lines show exact dynamics, blue dashed lines show model predictions,
and open orange circles show independent IBM measurements
at the indicated times.
Each experimental point is estimated from
$1{,}024$ measurements in the $X$ basis, with
error bars denoting one standard error
from finite-shot sampling.
}
    \label{fig:5}
\end{figure*}

Fig.~\ref{fig:5} compares the predicted $\langle X_5(t)\rangle$ with independent experimental measurements and exact dynamics for three distinct initial states at $h=0.2$ and $h=0.6$.
Specifically, we test a tilted product state $[\cos(\pi/8)|\uparrow\rangle+\sin(\pi/8)|\downarrow\rangle]^{\otimes 10}$ lying outside the Pauli-eigenstate training set in Fig.~\ref{fig:5}(a,d), the $X$-basis N\'eel state $|\rightarrow\leftarrow\rightarrow\leftarrow\cdots\rangle$ in Fig.~\ref{fig:5}(b,e), and the $X$-polarized state $|\rightarrow\rangle^{\otimes 10}$ in Fig.~\ref{fig:5}(c,f).
Across all six panels, the model predictions closely track both the independent IBM measurements and the exact trajectories over $t \in [0, 1]$, accurately reproducing the distinct relaxation profiles as well as the slower decay at the larger transverse field $h=0.6$.
Together, these trajectories confirm simultaneous generalization on quantum hardware across initial states, Hamiltonian parameters, and continuous evolution times.
Crucially, this agreement holds uniformly across the processor.
Evaluated on $900$ held-out random product-state samples, the prediction RMSE relative to the experimental measurements ranges from $0.0361$ to $0.0526$ across all ten qubits, with per-qubit benchmarks reported in Fig.~S2. 

This predictive accuracy is achieved with
experimental data that contain not only
finite-shot fluctuations but also hardware imperfections
in state preparation, gate execution, and readout.
Although these imperfections can induce systematic
departures from ideal unitary evolution,
our findings show that the proposed learning
strategy reliably captures actual device
dynamics well beyond the unbiased measurement noise
model assumed in our theoretical analysis.

\section{Discussion and Outlook}
\label{sec:discussion}

We have introduced a foundation model built on physical principles that learns quantum dynamics directly from noisy experimental measurements without requiring knowledge of the microscopic Hamiltonian.  Remarkably, from a single dataset collected on simple random product states, the model achieves provable generalization simultaneously along multiple axes, spanning unseen initial states including entangled states, Hamiltonian parameters, continuous evolution times, and an entire class of bounded-degree observables.
For finite-range static and smoothly time-dependent Hamiltonian families, we proved that learning a prescribed observable at a fixed Pauli-weight cutoff requires a sample complexity independent of system size and at most polynomial in the evolution horizon, with worst-case optimal scaling in key special cases.
Moreover, reusing the same classical-shadow snapshots to predict all bounded-degree observables simultaneously incurs only a logarithmic overhead in system size.
Numerical benchmarks on static and periodically driven transverse-field Ising chains, combined with hardware experiments on a superconducting quantum processor, validate these theoretical guarantees across both product and entangled initial states, including dynamical regimes governed by quantum coherence.

Our framework is neither tied to a specific source of training data nor restricted to a single global Hamiltonian. The training data may be obtained from classical simulations, digital quantum processors, or analog quantum simulators, provided that the platform supports the required state preparation and randomized measurement protocols. In particular, as demonstrated in our hardware experiment, the framework applies equally when continuous-time evolution is discretized into local quantum circuits via Trotterization, since finite-depth local circuits naturally preserve a strict causal light cone. Furthermore, because Lieb--Robinson bounds together with the Pauli weight truncation confine the evolved observable to a light cone region of radius $R \sim vT$ around its support, the prediction depends only on the Hamiltonian terms and the reduced initial state within this local neighborhood. A model trained on one system therefore applies, up to a controlled truncation error, to any other Hamiltonian that coincides with it within the light cone, irrespective of differences in total system size, boundary conditions, or interactions elsewhere in the system.

Several concrete directions can further extend our framework. Algebraic Lieb--Robinson bounds could generalize our guarantees to power-law long-range interactions relevant to trapped ions and Rydberg atom arrays, while generalized classical shadows, such as shallow-circuit, fermionic, or bosonic protocols, would broaden applicability to analog simulators like ultracold atoms in optical lattices. To push predictions to longer times in chaotic regimes, incorporating physical symmetries and conservation laws to prune the operator basis, or replacing the linear Chebyshev expansion with expressive nonlinear neural architectures that preserve physically grounded inductive biases, could substantially improve scalability. In particular, combining our locality-based principles with neural networks offers a promising way to unite the expressivity of deep learning foundation model with provable generalization guarantees.

\section{Acknowledge}
This work was supported in part by the Beijing Major Science and Technology Project under Contract No. Z251100008125036.

\bibliography{ref}
\end{document}